\documentstyle[12pt,overcite,procs]{article}
\def\lsim{\frac{<}{\sim}}
\def\gsim{\frac{>}{\sim}}
\begin{document}
\begin{Titlepage}
\Title
A SOLVABLE MODEL FOR
ANHARMONIC EVOLUTION OF COSMIC AXION OVERDENSITIES
\endTitle
\Author{KARL STROBL}
Department of Applied Maths and Theoretical Physics,
Cambridge University, Cambridge, UK
\endAuthor
\And
\Author{THOMAS J. WEILER}
Department of Physics \& Astronomy,
Vanderbilt University, Nashville, TN 37235, USA
\endAuthor
\endAuthors
%

\begin{Abstract}
We present analytic solutions to the spatially homogeneous axion field
equation, using a model potential which strongly resembles the $(1-\cos
\theta)$
standard anharmonic potential.
We find that the anharmonicity amplifies early universe axion density
\mbox{f}luctuations,
but only significantly so for relatively large initial misalignment angles.
Enhancements of $\sim$ (2,3,4,13) result for
$\theta_{\rm in}\sim (0.85,0.90,0.95,0.99)\times\pi$.
The large $\theta$ result agrees with Lyth\cite{Lyth},
and so validates his approximations.
\end{Abstract}
\end{Titlepage}
\vspace{0.5cm}

Axions are a consequence of the simplest solution to the strong
CP problem\cite{PQ},
and are the prototype candidate for cold dark matter\cite{KoTur}.
The complex scalar field $\psi_{PQ}$
has at temperatures above the QCD scale
an effective potential of the ``Mexican hat" shape.
The axion field $a(\vec{x},t)$
corresponds to the scaled phase $f_{\rm a}\,\theta$
of $\psi_{PQ}$.  At low temperatures
$T\lsim \Lambda_{\mbox{\scriptsize QCD}} \sim$ 1 GeV,
the Mexican hat potential is
tipped due to instanton effects\cite{GPY},
yielding
$V(\theta)\simeq (m_a f_a)^2 (1-\cos\theta)$.
Current algebra relates pion and axion (both pseudo--Goldstone bosons)
dynamics,
from which one derives $m_a f_a \sim m_{\pi} f_{\pi}$, and
$(m_{\rm a}/10^{-5}\mbox{eV})\simeq (f_{\rm a}/10^{12}\mbox{GeV})^{-1}$.
We present a solvable model
which closely mimics this instanton--induced potential,
and  obtain the time--evolution of fluctuations
in the axion field and energy density,
{\sl including anharmonic effects}.
Our effective potential consists of a parabola matched to an inverted parabola:
$V(\theta)= \frac{1}{2}f_{\rm a}^2 m_{\rm a}^{2}(T)\theta^2$
for $\left|\theta\right|	<\frac{\pi}{2}$, and
$= \frac{1}{2}f_{\rm a}^2 m_{\rm a}^{2}(T)\left(\frac{\pi^2}{2}-
     \left(\pi-\left|\theta\right|\right)^2\right)$
for $\left|\theta\right|>\frac{\pi}{2}$.
As shown in Fig. (1),
this functional form strongly resembles the familiar cosine shape and
preserves the periodicity;
at the same time it yields a solvable homogeneous equation of motion (eom).

\begin{figure}

\begin{center}
		   \fbox{\vbox{\hrule height0pt depth0pt width3in \vskip2in
                 \hrule height0pt depth0pt width3in}}
\end{center}
\caption{
A comparison of our model potential (solid line)
with the cosine potential (dashed line); and a
schematic drawing of the field oscillations evolving from
a large misalignment angle, overshooting the convex part of the
potential once.  The height of the potential is a function
of time, or equivalently, temperature; the shape is not.}
\label{fig:one}
\end{figure}

In a radiation-dominated FRW universe, the eom for the homogeneous ``zero
mode'' axion field $a(t)=f_a \theta(t)$ is
$\ddot{\theta}(t)+\frac{3}{2t}\dot{\theta}(t)+m_{\rm a}^2(T)\theta(t)=0$,
if $\left|\theta\right|\leq\pi/2$, and the same equation with $\theta$ replaced
by $\epsilon=\pm\pi-\theta$
and $m_{\rm a}^2$ by $-m_{\rm a}^2$ for $|\theta|>\pi/2$.
We neglect gradient terms;
for small fluctuations, such as those predicted in inflationary models,
this introduces negligible errors.
Since the equation of motion is linear and homogeneous in $\theta$ for
$\left|\theta\right| \leq\pi/2$, the mean field $\overline{\theta}$
and the local field $\theta=\overline{\theta}+\delta\theta$ evolve with
the same time--dependence. Thus, the field contrast $\frac{\delta\theta}
{\overline{\theta}}$ is time--independent, i.e. does not evolve.
The same conclusion follows for $\frac{\delta\epsilon}{\overline{\epsilon}}$
at  $|\theta|>\pi/2$.  Growth cannot occur while the field remains in a
harmonic region.
In our model, then, the evolutionary enhancement of
$\frac{\delta\theta}{\overline{\theta}}$ must
come from the matching of solutions at $\theta=\pm\pi/2$.
Growth occurs because $\theta=\overline{\theta}+\delta\theta$ must be matched
to $\pm\pi/2$ at a time differing slightly from the matching time of
$\overline{\theta}$.

The only significant
growth in density perturbations occurs at the time when the
growing axion mass is comparable in magnitude to the diminishing
Hubble parameter.
At earlier times the fluctuations are frozen by the overdamped equation of
motion; and for later times, as we will show,
the field has evolved into the harmonic region of the potential where growth
is not possible.

A power law $m_{\rm a}\propto T^{-n}$ turn--on of the axion mass (i.e.
potential's curvature) is seen in perturbative instanton calculations;
we assume this form here. The early universe
relation\cite{KoTur} between time and temperature is
$t^{-1}=26 \frac{T^2}{M_P}$.
At $T\!\sim1$~GeV, the value of $3H \sim 3\times10^{-9}$~eV is
already much smaller than the zero--temperature axion mass\cite{KoTur}
($\gsim 10^{-5}$ eV).
Thus, once the instanton--induced potential
turns on at $T\sim1$~GeV, its curvature quickly dominates the Hubble
term in the equation of motion.
Ultimately, the axion mass reaches its
zero--temperature value and ceases to grow.

It is convenient
to measure all time in units of $\tilde{t}$,
the time (roughly) when the axion oscillations
for the convex part of the potential first become undamped.
$\tilde{t}$ is defined implicitly by
$m_{\rm a}(\tilde{t})=\frac{3}{2\tilde{t}}$,
which yields $\tilde{T}\sim 1$ GeV and $\tilde{t}\sim 10^{-6}$ sec.
In terms of this scaled, dimensionless time, the field equation is simply
$\ddot{\chi}(t)+\frac{3}{2t}\dot{\chi}(t)\pm\frac{9}{4}t^n\chi(t)=0$, where in
the convex part of the effective potential $\chi$ denotes $\theta$ and the
plus sign holds for the last term;
in the concave part of the potential,
$\chi$ denotes $\epsilon=\pm\pi-\theta$ and the minus sign holds.
The solutions
are the Bessel and modified Bessel functions, respectively:\\
$\theta(t)  =  t^{-\frac{1}{4}}J_{\pm\frac{1}{2n+4}}\left(
\frac{3\,t^{1+n/2}}{n+2}\right)$ and
$\epsilon(t)  =  t^{-\frac{1}{4}}I_{\pm\frac{1}{2n+4}}\left(
\frac{3\,t^{1+n/2}}{n+2}\right)$.

The initial field at $|\theta_{\rm in}|>\frac{\pi}{2}$, i.e.
$\epsilon_0(t_{\rm o}\!\ll 1)=\epsilon_{\rm in}$, is given by
the modified Bessel with positive index,
since the negative index solution is singular at $t=0$.
Then each time the
field amplitude passes through $\pm\pi/2$
from above or below, the $\theta$ and $\epsilon$ solutions, and their first
derivatives, must be matched.
The final field amplitude
is generated in this recursive way.
It is useful to introduce still another rescaling of time,
$z\equiv\frac{3\,t^{\frac{n+2}{2}}}{n+2}$,
as given in the arguments of the Bessel functions.
We label the matching times $z_i$, and display them in Fig.\ (2).

\begin{figure}

\begin{center}
		   \fbox{\vbox{\hrule height0pt depth 0pt width 3in \vskip2in
                 \hrule height0pt depth 0pt width 3in}}
\end{center}
\caption{
Matching times $z_1$ through $z_5$ obtained from the transcendental equations,
vs.\ $\overline{\epsilon_{{\rm in}}}$.}
\label{fig:two}
\end{figure}

It is seen that the values of the $z_i$ hardly depend on $n$,
whereas $t_i=((n+2)z_i/3)^{\frac{2}{n+2}}$ depend sensitively
on $n$.
Several important inferences may be drawn from Fig.\ (2):
(i) the convex region is overshot (for $n=3.7$)
when $\epsilon_{\rm in}\stackrel{<}{\sim}0.02\,\pi$, and doubly overshot
when $\epsilon_{\rm in}\stackrel{<}{\sim}10^{-3}\,\pi$;
(ii) the $z$--time spent in the overshot concave region,
$z_3-z_2$, is relatively short;
(iii) we do not need to cut off the power--law growth of mass in our
calculations, since the anharmonic effects which govern density
growth terminate before the axion mass reaches its
low--temperature value:
with the power--law growth
the mass saturates at a $z$ value of
$z_c=(\frac{3}{n+2})(m_{\rm a}(\infty)/m_{\rm a}(\tilde{t}))^{\frac{n+2}{n}}
\stackrel{>}{\sim} 10^{4}$,
much larger than the occurrences of
the matching times $z_1$ through $z_5$, even for infinitesimal values
of $\epsilon_{\rm in}$.

To quantify the final axion density contrast,
$\frac{\delta{\rho_{\rm a}}}{\overline{\rho_{\rm a}}}$,
we find it convenient to define its dependence on
the initial field misalignment by
$\frac{\delta \rho_{\rm a}}{\overline{\rho_{\rm a}}}=
\frac{1}{2}\xi(\overline{\theta_{\rm in}})
\frac{\delta\theta_{\rm in}^2}{\overline{\theta_{\rm in}^2}}
\approx \xi(\overline{\theta_{\rm in}})
\frac{\delta\theta_{\rm in}}{\overline{\theta_{\rm in}}}$.
Since $\frac{\delta\rho}{\overline{\rho}}$ is just
$\frac{\delta\theta_{\rm out}^2}{\overline{\theta_{\rm out}^2}}$,
we may also write
$\xi(\overline{\theta_{\rm in}}) \approx 2\frac{\overline{\theta_{\rm in}}}
{\overline{\theta_{\rm out}}}\frac{d\theta_{\rm out}}{d\theta_{\rm in}} =
2\frac{\overline{\theta_{\rm in}}}{\overline{\epsilon_{\rm in}}}
\left(-\frac{d\;\ln(\theta_{\rm out})}{d\;\ln(\epsilon_{\rm in})}\right)$.
In the harmonic regime ($\theta_{\rm in}<\frac{\pi}{2}$ in our model),
$\rho_{\rm a} \propto \theta^2_{\rm in}$,
and so $\xi=2$. For $|\theta_{\rm in}|>\frac{\pi}{2}$,
we expect anharmonic effects to retard the
progression of the amplitude toward the minimum of the potential,
thereby effecting a larger density contrast, $\xi(\theta_{\rm in})>2$.
In Fig.\ (3)
we show the exact $\xi$ obtained from our model for different values of $n$.
The enhancement, $\xi/2$, is large for small $\epsilon_{\rm in}$;
e.g. it is a factor
of $\sim$ (2,3,4,13) for $\epsilon_{\rm in}\sim
(0.15,0.10,0.05,0.01)\times\pi$.
This is our most important result.

\begin{figure}

\begin{center}
   \fbox{\vbox{\hrule height0pt depth 0pt width 3in \vskip2in
                            \hrule height0pt depth 0pt width 3in }}
\end{center}
\caption{
$\xi(\overline{\theta_{\rm in}})$ versus
$\overline{\epsilon_{\rm in}}$.
$\xi$ is bounded from below by the no--enhancement, harmonic value
$\xi\equiv 2$.}
\label{fig:three}
\end{figure}

The occurrence of the large density fluctuations at large $\theta$
is possibly impacted by density--fluctuation bounds,
e.g.\ COBE measurements.
The heuristic result of ref.(1)
is in good agreement with our result at large $\theta$.
Thus, our exclusions in the $m_{\rm a}$--$\theta_{\rm in}$ plane
are the same as those of Lyth \cite{Lyth}.
Given a model for the initial field fluctuation
spectrum $\frac{\delta\theta_{\rm in}}{\overline{\theta_{\rm in}}}$,
these results may exclude particular models.
Our evolution algorithm, which avoids the harmonic and adiabatic
approximations,
is ready and able to turn an initial fluctuation
spectrum satisfying $\delta\theta/\overline{\theta} \ll 1$ initially,
into a final density contrast;
the algorithm is valid over the entire region
$\theta_{\rm in} \in [-\pi,\pi]$.
Further details and additional aspects of this model are available\cite{sw}.

\section*{Acknowledgements}

This work is supported in part by U. S. DOE grant DE--FG05--85ER40226,
the Austrian Ministry of Science, the Fulbright Commission, and the Cambridge
Overseas Trust.



\end{document}